\documentclass{article}
\usepackage{epsfig}
\usepackage{amssymb}
\usepackage{amsmath} 
\begin{document}

\title{THE GROUND STATE STRUCTURE OF ELECTRON'S ENSEMBLE ON ONE-DIMENSION DISORDERED LATTICE}
\author{L.A.Pastur, V.V.Slavin, A.A.Krivchikov \\
B.I Verkin Institute for Low-Temperature Physics and Engineering\\
of National Academy of Sciences of Ukraine.\\
47 Lenin Ave., Kharkov, 61103, Ukraine.\\
e-mail:slavin@ilt.kharkov.ua}

\maketitle

\begin{abstract}
The ground state of interacting particles on  a disordered one-dimensional
host-lattice is studied by a direct numerical method. It is shown that if
the concentration of particles is small, then even a weak disorder of the
host-lattice breaks the long-range order of Generalized Wigner Crystal,
replacing it by the sequence of blocks (domains) of particles with random
lengths. The mean domains length as a function of the host-lattice disorder
parameter is also found. It is shown that the domain structure can be
detected by a weak random field, whose form is similar to that of the ground
state but has fluctuating domain walls positions. This is because the
generalized magnetization corresponding to the field has a sufficiently
sharp peak as a function of the amplitude of fluctuations for small
amplitudes.
\end{abstract}

\section{Introduction}
Low-dimensional and layered conductors possess a number of interesting and
unusual properties and have been of considerable interest in the last
decades. In particular, it is possible to separate electron subsystem and
that of dopant ions in a number of these conductors. As a result, the
potential, produced by the dopants in the conducting electron layers is
almost constant and conducting characteristics are essentially determined
by the inter-electron (or inter-holes) repulsion.

Another interesting subclass of these conductors consists of lattice electrons
systems, where the charge carriers tunneling between host-lattice sites is
suppressed by their mutual Coulomb repulsion. As a result, the charge
carriers are \emph{self-localized} \cite{Mo:74,Slutskin_Gorelik}.These are
the MOSFETs (metal-oxide-semiconductor field-effect transistors) and a
variety of other semiconducting heterostructures. The quasi-one-dimensional
organic conductors \cite{Hubbard1} and artificial systems like arrays of
quantum dots \cite{nano_dot}, networks and chains of metallic nano-grains
with tunnel junctions (provided by various organic molecules)
\cite{nano_grain} also belong to this class.

The ground state (GS) and the low-temperature properties of these systems
have been extensively studied. The first theoretical result is due to
Hubbard \cite{Hubbard1}, who considered the GS of one-dimensional (1D)
repulsing particles on a periodic host-lattice in the limit of strong
interaction, where the dynamic effects are negligible and particles are
localized with high accuracy on the lost-lattices sites. It was shown in
\cite{Hubbard1} that the corresponding GS has, in general, an
incommensurable structure determined only by the particle density
\begin{equation}
c_{e}=N_{e}/L,  \label{ce}
\end{equation}
where $N_{e}\gg 1$ and $L\gg 1$ are the total numbers of particles and host lattice sites, but does not depend on pair potential $V$, provided that $V$ is
non-negative, convex and decays at infinity faster then $|x|^{-1}$.

Hubbard also suggested an explicit form of the GS in this case. The form was
justified and clarified in \cite{Bak,Sinay} and it is known now as the
Generalized Wigner crystals\textquotedblright\ (GWL). According to the GWL
theory, the position $x_{k}$ of $k$-th particle is given by the
simple formula:

\begin{equation}
x_{k}=a_{0}[k/c_{e}+\phi ],.k=1,...,N_{e},  \label{x_k}
\end{equation}
where $[\ldots ]$ denotes the integral part, $\phi $ is an arbitrary
constant (initial) phase and $a_{0}$ is the distance between the host
lattice sites. It follows from the formula that the only two inter-particle
distances
$$x_{k+1}-x_{k}=[1/c_{e}],\;[1/c_{e}+1]$$

can appear in the GS, depending on $\phi$ and $c_{e}$ \cite{Hubbard1}.
Thus, the inter-particle distances do not coincides in general with the
minima of potential energy of the system. This leads, in particular, to
rather unusual zero-temperature dependence of $c_{e}$ on the chemical potential
$\mu $, which proved to be fractal (devil staircase) \cite{Sinay}.

The low-temperature thermodynamics of 1D GWL was considered in \cite{PRB96}.
In \cite{PRB00} it was shown that 2D systems are characterized by an effective
reduction of the dimension that allows one to find their GS, which proved to
be a natural 2D generalization of 1D Hubbard description (see (\ref{x_k})).

Note, however, that the majority of 1D self-localized conductors are
disordered. For example, in semiconductors on the base of MOSFET the
disorder is due random impurities, in nanostructures
\cite{nano_dot,nano_grain} the disorder is due to the fluctuations of tunneling
junctions, and in the 1D charge transfer salts \cite{Hubbard1} the disorder
results from imperfections of their chemical structure. That is why the
influence of host-lattice disorder on the GS properties is of great
interest, especially taking into account that in the 1D case even a weak
disorder of the host-lattice can affects essentially the GS structure and
low temperature thermodynamic properties. It suffices to recall the
Larkin-Imry-Ma result \cite{Larkin,Imry-Ma}, according to which an arbitrary
weak random field destroys the long-range order in 1D and 2D systems.

As was mentioned above, the systems under consideration have been
extensively studied both theoretically and experimentally. However, the
majority of theoretical results are obtained in the frameworks of rather
simple models, providing often only qualitative estimates.

In this situation is it natural to use numerical methods of analysis of thermodynamic properties of 1D systems, since in the 1D case rather large systems (up to $N_e\sim 10^4-10^5$ particles) can by studied  by the transfer-matrix method. However, the use of the method to find the the low temperature asymptotic regime is rather limited by the exponential growth (or decay) of the entries of transfer matrix. 
This implies the considerable decrease of calculation accuracy or even an overflow in numerical co-processor registers.

In this paper we propose a new numeric method for the study of GS of the system in question, i.e., classical 1D repulsing particle on a disordered
host-lattice in the limit of low particle density (\ref{ce}).
This condition is the case for many 1D lattice systems,
including quasi-one-dimensional organic conductors, chains of nano-objects,
etc.

\section{Model}
We will describe the one-dimensional system of repulsing particles
\cite{Hubbard1,Frattini1,Frattini2} on disordered host-lattice by the extended
Hubbard model determined by the Hamiltonian \cite{Leib}

\begin{equation}
{\hat{H}}=-t\sum\limits_{<\alpha, \beta>,\sigma }{\hat{c}}
_{R_{\alpha },\sigma }^{+}{\hat{c}}_{R_\beta,\sigma }+U\sum\limits_{\alpha}{\hat{n}}
_{R_{\alpha },\uparrow }{\hat{n}}_{R_{\beta },\downarrow }+\frac{1}{2}%
\sum\limits_{\alpha\neq \beta}V(|R_{\alpha }-R_{\beta }|){\hat{n}}_{R_{\alpha }}{\hat{
n}}_{R_{\beta }}.  
\label{Hub1}
\end{equation}
where $<\alpha,\beta>$ means summation over the nearest neighbors, $\{R_{\alpha }\}$ are the sites of host-lattice ($\alpha =1,2,\ldots ,L
$), ${\hat{c}}_{R_{\alpha },\sigma }^{+}$ and ${\hat{c}}_{R_{\alpha
},\sigma }$ are the creation and annihilation operators on the site $R_{\alpha
}$ with spin $\sigma $ ($\sigma =\uparrow ,\downarrow $), $U$ is the
interaction energy of two particles on the same site of the host-lattice,
${\hat{n}}_{R_{\alpha },\sigma }={\hat{c}}_{R_{\alpha },\sigma }^{+}{\hat{c}}
_{R_{\alpha },\sigma }$ is the number operator with a fixed spin, ${\hat{n}}
_{R_{\alpha }}={\hat{n}}_{R_{\alpha },\uparrow }+{\hat{n}}_{R_{\alpha
},\downarrow }$ is the total number operator at $R_{\alpha }$.

We will consider systems for which the potential has the power-low decay at
infinity, i.e.,
\begin{equation}
V(|x|)\sim |x|^{-\gamma },\;\;\gamma >0,\quad |x|>>a_{0}.
\label{Coulomb_stability}
\end{equation}
where $a_{0}$ is the length fixing the order of magnitude the distance
between the adjacent host-lattice sites, i.e., is the period of translation
invariant lattice, the mean distance between the sites of a disordered
lattice, etc. (cf. (\ref{x_k})).

The thermodynamic stability requires that $\gamma >1$ \cite{Sinay}. Thus, we
write
\begin{equation}
\gamma =1+\delta, \; \delta>0
\label{gls}
\end{equation}
and assume that $V$ of (\ref{Coulomb_stability})
is positive (repulsive) and convex. The last condition is important and
its violation may change significantly the results 
(see e.g. \cite{Jedrzejewski}).

Note that the systems for which $-1<\delta\leq 0$ has also been studied
recently (see e.g. \cite{Ca-Co:08,Da-Co:09,Mu-Co:10}). One can call these
systems the ``strong'' long ranged. In this
case one has to either take into account certain truncations procedures
(screening, confining the system to a finite box, etc.) or to be prepared to
obtain rather unusual properties, especially if $V$ is attractive or
anisotropic \cite{Ca-Co:08,Da-Co:09}. On the other hand, the
systems where with $\delta>0$) can be called the ``weak'' long ranged.
They admit the traditional statistical
mechanics description, possessing, however, certain special properties if 
$\delta $ is small (say, $0<\delta \leq 2$).

We will consider in this paper the limit of low particle density $c_{e}$ of 
(\ref{ce})
\begin{equation}
c_{e}=N_{e}/L<<1.
\label{cesm}
\end{equation}
Besides, we will neglect both the dynamic effects ($t=0$) and the effects of
Fermi statistics ($U\rightarrow \infty$) \cite{Hubbard1}. The last limit
leads to so called spinless fermions model. As a result, the only potential
energy of particle repulsion has to be taken into account and the
Hamiltonian (\ref{Hub1}) can be replaced by

\begin{equation}
H=\frac{1}{2}\sum\limits_{\alpha \neq \beta }V(|R_{\alpha }-R_{\beta
}|)n_{R_{\alpha }}n_{R_{\beta }},  
\label{Hub2}
\end{equation}
where $\{n_{R_{a}}\}$ are the classical occupation numbers: $n_{R_{\alpha
}}=0,1$.

It is convenient to pass from the occupation numbers $\{n_{R_{a}}\}$ of the
host-lattice sites $\{R_{\alpha }\}$ to the coordinates $\{x_{k}\}$ of
particles ($k=1,...,N_{e}$), i.e., the coordinates of points of the
host-lattice where the occupation numbers equal 1. We obtain instead of 
(\ref{Hub2})
\begin{equation}
H=\sum\limits_{j<k}V(|x_{j}-x_{k}|).  \label{Hx}
\end{equation}
It was shown in \cite{PSS1} that in the low-temperature limit the major
contribution into partition function of (\ref{Hub2}) is due to the particle
configurations which are close to equidistant ones, i.e., to the ground
state configuration of the Wigner crystal with the same $c_{e}$. In other
words, if $l_{0}=a_{0}/c_{e}$ is the period of the Wigner crystal, hence the
coordinate $r_{k}$ of its $k$-th site is $r_{k}=kl_{0}$, and if $x_{k}$ is
the coordinate of $k$-th site of Generalized Wigner Crystal, then the
typical shifts between $r_{k}$ and $x_{k}$ are
\begin{equation}
\overline{s}=<|x_{k}-r_{k}|>\sim a_{0}<<l_{0},  \label{tysh}
\end{equation}
where the symbol $<\ldots >$ denotes the averaging with respect to the
host-lattice disorder which we assume as usually to be translation invariant
in the mean \cite{LGP:88}

We denote the shifts
\begin{equation}
s_{k}=x_{k}-r_{k}  \label{sal}
\end{equation}
and expand the Hamiltonian (\ref{Hub2}) with respect to the small parameter 
$\overline{s}/l_{0}\simeq a_{0}/l_{0}<<1$:

\begin{equation}
H\approx H_{WC}(c_{e})+\frac{1}{2}\sum\limits_{\overset{j,k=1}{j\neq k}
}^{N_{e}}b_{j-k}(s_{j}-s_{k})^{2}.  
\label{Hub3}
\end{equation}
Here $H_{WC}$ is a constant (the ground state energy of Wigner crystal with
a given density) and
$$
b_{j-k}=b(|r_{j}-r_{k}|)=b(|j-k|l_{0})=\frac{1}{2}\frac{\partial ^{2}V(x)}
{\partial x^{2}}|_{x=|j-k|l_{0}}.$$

\begin{figure}
\centerline{\psfig{file=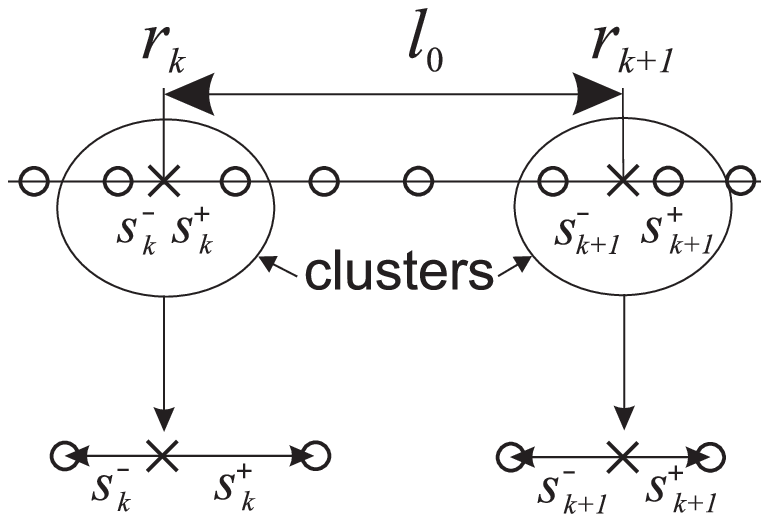, width=6.0cm}}
\vspace*{8pt}
\caption{1D disordered host-lattice (empty circles) and clusters.
The crosses are the sites of Wigner lattice with the same particle density $%
c_e$. The bottom shows the same clusters in larger scale.}
\label{Fig1}
\end{figure}

Omitting $H_{WC}$ and using the nearest neighbor approximation, we obtain
from (\ref{Hub3})
\begin{equation}
H=b_{1}\sum\limits_{\alpha =1}^{N_{e}-1}(s_{\alpha }-s_{\alpha +1})^{2}.
\label{Hub4}
\end{equation}
Note that in the above nearest neighbor approximation we take into account
the interaction between \textit{the nearest particles} but not the nearest
host-lattice sites. Since the typical distance between particles is 
$l_{0}=a_{0}/c_{e}$, hence $l_{0}>>a_{0}$ (see (\ref{tysh})) in the low
concentration limit (\ref{cesm}), the approximation seems fairly reasonable.

It is convenient to measure the energy in units of $b_{1}a_{0}^{2}$. Then 
(\ref{Hub4}) became

\begin{equation}
H=\sum\limits_{k=1}^{N_{e}-1}(s_{k}-s_{k+1})^{2},  
\label{Hub5}
\end{equation}
where now $H$ and the dynamic variables $\{s_{k}\}$ are dimensionless.
It was shown in \cite{PSS1} that if the temperature is low enough,
then it suffices to consider the case where $k$-th particle occupies only one
of \textit{two} host-lattice sites adjacent to the $k$-th site $r_{k}$ of
the corresponding Wigner crystal (see Fig.~\ref{Fig1}).

We will call these sites \textit{clusters}, each cluster\ consists of two
sites of the host lattice and contains only one particle which can occupy
one of these two host-lattice sites. In this case the shifts (\ref{sal}) can
be written as:

\begin{equation}
s_{k}=\sigma _{k}\lambda _{k}^{\sigma _{k}},\;\lambda _{k}^{\sigma }\geq 0,
\label{sapm}
\end{equation}
where $\{\sigma _{k}=\pm 1\}$ are the standard Ising spins and $\{\lambda
_{k}^{\sigma }\}$ are the random distances of the cluster sites from the
site $r_{k}$ of the Wigner crystal of the same density. We will assume that
the distances $\{\lambda_k^{\sigma }\}$ are independent and identically
distributed for all $k=1,2,\ldots ,N_{e}$, hence the
left hand and the right hand distances $\lambda _{k}^{+}$ and
$\lambda_k^{-}$ are typically different.

It follows from (\ref{sapm}) that the shifts (\ref{sal}) can
be viewed as Ising-type spins with random lengths $\lambda _{k}^{\sigma
_{k}},\;k=1,...,N_{e}$. To make more explicit the dependence of Hamiltonian
(\ref{Hub5}) on the dynamic variables $\{\sigma _{k}\}$ and the frozen
disorder described by the random spin lengths $\{\lambda _{k}^{\sigma _{k}}\}
$, we write
\begin{equation}
\lambda _{k}^{\sigma _{k}}=\alpha _{k}\sigma _{k}+\beta _{k},\;\alpha
_{k}=(\lambda _{k}^{+}-\lambda _{k}^{-})/2,\;\beta _{k}=(\lambda
_{k}^{+}+\lambda _{k}^{-})/2\geq 0,  \label{lab}
\end{equation}
i.e., $\{\alpha_k\}$ and $\{\beta_k\}$ do not depend on
$\{\sigma_k\}$ and are independent for different $k$. By using this
parametrization, we can rewrite (\ref{Hub5}) as

\begin{equation}
H(\{\sigma_k\})=C-2\sum_k\beta_k\beta_{k+1}\sigma_k\sigma
_{k+1}-\sum_k h_k\sigma_k
\label{HIs}
\end{equation}
where $C$ does not depend on $\{\sigma _{k}\}$ and can be omitted, $\{\beta_k\}$
are given by (\ref{lab}) and
\begin{equation}
h_k=2\beta_k(\alpha_{k+1}+\alpha_{k-1}-2\alpha_{k}).
\label{hk}
\end{equation}
Thus, the Hamiltonian (\ref{Hub5}) is thermodynamically equivalent to that
(\ref{HIs}) of the one-dimensional Ising model with random interaction and
random short correlated external field (recall that $\{\alpha_k\}$ and
$\{\beta_k\}$ are independent for different $k$). Since
$\{\lambda_k^{\sigma }\}$ are non-negative, independent and identically
distributed for all $k=1,2,\ldots, N_{e}$ and $\sigma =\pm 1$, it follows
from (\ref{lab}) that $\{\beta_k\}$ are non-negative, independent and
identically distributed and $\{h_k\}$ are symmetrically distributed, in
particular
\begin{equation}
<h_{k}>=0.  \label{mhk}
\end{equation}
We obtain the periodic host-lattice setting putting $\lambda_k^{\sigma
_k}=1$ for all $k$. In this case (\ref{HIs}) corresponds to the
one-dimensional ferromagnetic Ising model with the nearest-neighbor
interaction.

In general, each realizations of $2N_{e}$ random lengths
$\{\lambda_k^{\sigma_k}\}$, hence random variables $\{\beta_k\}$ and $\{h_k\}
$ provides a realization of disorder, thus fixing the Hamiltonian to be used
to find the partition function
\begin{equation}
Z(L,N_{e})=\sum_{\{\sigma _{k}=\pm \}}e^{H/T}  \label{pf}
\end{equation}
of our disordered system or its ground state $\{\sigma _{k}^{GS}\}$:
\begin{equation}
E_{GS}=\min_{\{\sigma _{k}=\pm \}}H(\{\sigma _{k}\})=H(\{\sigma _{k}^{GS}\}).
\label{gs}
\end{equation}

\section{The Ground State.}

The low temperature thermodynamic properties of the model (\ref{Hub5}) have
been studied in \cite{PSS1} using the transfer-matrix formalism. In
particular,  a weak local external field
\begin{equation}
h=\{\varepsilon h_{k}\},\;h_{k}=\delta _{k,k_{0}}  \label{hloc}
\end{equation}
was introduced into the Hamiltonian (\ref{Hub5}) as a tool of analysis of
the ground state. Here $\varepsilon$ is a small constant and $\delta
_{k,k_0}$ is Kronecker symbol. In other words, the field affects only the
$k_0$-th spin and the corresponding Hamiltonian is
$$
H=\sum\limits_{k=1}^{N}(s_{k}-s_{k+1})^{2}-\varepsilon s_{k_{0}}.$$

Using the transfer-matrix techniques, one can calculate the free energy
$F(T,h)$ of the system and the corresponding magnetization per spin
\begin{equation}
M_{loc}(T,k_{0})=-\frac{1}{N_{e}}\left. \frac{\partial F(T,h)}{
\partial \varepsilon }\right\vert _{\varepsilon =0}\approx \frac{
F(T,0)-F(T,h)}{N_{e}\varepsilon }.  \label{Mloc}
\end{equation}
It is reasonably to believe that if $M_{loc}(T,k_{0})>0$ in the low temperature limit,
then the $k_{0}$-th spin of the ground state is parallel to
the field, while if $M_{loc}(T,k_{0})<0$, then the $k_{0}$-th spin is
antiparallel to the field. Thus, calculating $M_{loc}(T,k_{0})$ for
$k_{0}=1,2,\ldots ,N_{e}$ one can obtain the orientations of all the spins of
ground state.

However, the method seems to have certain disadvantages. First, since $h$
affects only \textit{one} spin, the local magnetization (\ref{Mloc}) is of
the order $1/N_{e}<<1$ for sufficiently large systems. Hence, the method can
only be applied to the cases, where the length of the system is not too
large, thus the boundary conditions may seriously effect the ground state.
Second, the amplitude $\varepsilon $ of the local field  should be small:
\begin{equation}
\varepsilon << T,e_{sf},  \label{e_sf}
\end{equation}%
where $e_{sf}$ is typical ``spin flip''
energy. However, in view of possible local energy degeneration $e_{sf}$ can
be zero for certain spins, thus the sign of $M(T,k_{0})$ is rather sensitive
to the value of $\varepsilon$ and the applicability of the method to rather
large systems is again questionable.

In this paper we study the ground state by a new method, which is free from
the above disadvantage. The main idea of the method is as follows. Let us
divide the system into $n$ parts (subchains) $C_{m},\;m=1,...,n$ with the
endpoints $p_{0},p_{2},\ldots ,p_{n}$, where $p_{0}=1$, $p_{n}=N_{e}$ and
$$
p_{0}<p_{1}<p_{2}<\ldots <p_{n-1}<p_{n}.$$

The lengths of the $m$-th subchain is $l_{m}=p_{m+1}-p_{m}+1$,$
\;m=0,1,\ldots ,n-1$). Thus, the $m$-th subchain contans $l_{m}$ spins

$$\sigma _{p_{m}},\;\sigma _{p_{m}+1},\ldots ,\sigma _{p_{m+1}-1},\;\sigma
_{p_{m+1}},$$
and the spin $\sigma _{p_{m}}$ ($m=1,2,...,n-1$) belongs to the \textit{two}
neighboring subchains, i.e., $\sigma _{p_{m}}$ is the last
``spin'' of $m$-th subchain and the first spin of $(m+1)$-th subchain.

It is convenient to index the spins in each subchain as
$$\sigma _{p_{m}+j-1}=\sigma _{m,j},$$

i.e., the first index indicates that spin belongs to the $m$-th subchain and the
second one is the number of the spin in the subchain. In this notation the 
$m$-th subchains consists of the spins
$$\sigma _{m,1},\sigma_{m,2},\ldots ,\sigma _{m,l_{m}-1},\sigma _{m,l_{m}}.$$
Now we carry out the direct enumeration of the states (the direct search for
the configurations with minimum energy) in each of $n$ subchains.

According to (\ref{Hub5}), the energy of $m$-th subchain is:
\begin{eqnarray*}
H_{m}&=&H_{m}(\sigma _{m,1},\sigma _{m,2},\ldots ,\sigma _{m,l_{m}}) \\
&=&-2\sum_{j=1}^{l_{m}-1}\beta _{m,j}\beta _{m,j+1}\sigma _{m,j}\sigma
_{m,j+1}-\sum_{j=1}^{l_m-1}h_{m,j}\sigma _{m,j}
\end{eqnarray*}
Let $H_{m}^{\min }$ is the minimum of $H_{m}$ over the spin configurations
of the subchain and let
\begin{equation}
\sigma _{m,1}^{\min },\sigma _{m,2}^{\min },\ldots ,\sigma
_{m,l_{m}-1}^{\min },\sigma _{m,l_{m}}^{\min }.  \label{min_chain}
\end{equation}
be a corresponding spin configuration: $H_{m}^{\min }=H_{m}(\sigma
_{m,1}^{\min },\ldots ,\sigma _{m,l_{m}}^{\min })$.

Now we note that if the last spin of each subchain is equal to the first
spin of the next subchain, i.e., if
\begin{equation}
\sigma _{m,l_{m}}^{\min }=\sigma _{m+1,1}^{\min },\;m=1,2\ldots ,n-1,
\label{shivka}
\end{equation}
then the union of all the subchain minimizing configurations (\ref{min_chain})
is a ground state configuration $\sigma^{GS}=\{\sigma_k^{GS}\}$ of the
whole Hamiltonian (\ref{HIs}):
\begin{eqnarray*}
\sigma ^{GS} &=&(\sigma _{1}^{GS},\sigma _{2}^{GS},\ldots ,\sigma
_{n-1}^{GS}) \\
&=&(\sigma _{0,1}^{min},\sigma _{0,2}^{min},\ldots ,\sigma
_{0,l_{0}}^{min},\sigma _{1,2}^{min},\ldots ,\sigma _{1,l_{1}}^{min},\ldots
,\sigma _{n-1,2}^{min},\ldots ,\sigma _{n-1,l_{n-1}}^{min}),
\end{eqnarray*}
and the corresponding total minimum energy (\ref{gs}) is
\begin{equation}
E_{GS}=\sum\limits_{m=0}^{n-1}H_{m}^{\min }.  
\label{method_restriction}
\end{equation}
A similar idea has been recently used to study ground states of certain
rather complex (in particular frustrated) translation invariant spin models
\cite{Gr:12,Du:12}.

The above suggests a direct numerical algorithm to search ground states:
split the system into subchains, minimize the energy of every subchain and
check the matching conditions (\ref{shivka}). If the conditions are not
satisfied, repeat the procedure.

Note, however, that the procedure does not guarantee that the ground state
is unique.

The choice of an optimal number $n$ of subchains depends on the
computer efficiency. Since in this scheme we perform the direct enumeration
of the states (direct energy minimization) for each ``subchain'',
the typical calculation time is $t_{0} \sim n2^{N_{e}/n}$. It is thus reasonable to choose $n$ so that $t_{0}$ is
several seconds, i.e., $n\sim N_{e}/10-N_{e}/20$, where $N_e = 10^4 - 10^5$
and even more. Increase in $n$ leads to
decrease in enumeration time, but at the same time, to increase in the number
of attempts (number of generations of division points $\{p_{m}\}$).

It should be noted that the proposed method is rather universal and can be
applied for a wide class of 1D disordered systems. An important advantage of
the methods is that the direct minimization of energy is carried out
independently in each subchain, thus the corresponding operations can be
easily adapted to the parallel calculations. Besides, the method can be modified
to deal with systems with larger number of interaction neighbors. In this
case the conditions (\ref{shivka}) is modified. For example, if we take into
account the near- and next-neighbors interaction, then the matching
conditions (\ref{shivka}) are
$$
\left\{
\begin{array}{l}
\sigma _{m,l_{m}-1}^{\min }=\sigma _{m+1,1}^{\min } \\
\sigma _{m,l_{m}}^{\min }=\sigma _{m+1,2}^{\min }.
\end{array}
\right.
$$
By using the method, we studied the ground state of the Hamiltonian
(\ref{HIs}). It is convenient to quantify the amount of disorder by writing the
random spin length (\ref{sapm}) as%
\begin{equation}
\lambda _{k}^{\sigma }=1-\xi _{k}^{\sigma },\;\xi _{k}^{\sigma }=2A\eta
_{k}^{\sigma },  \label{s_k1}
\end{equation}
where $\{\eta _{k}^{\sigma }\}$ are independent random variables uniformly
distributed over the interval $[0,1)$ ($<\eta _{k}^{\sigma }>=1/2$) and $A$ is
the ``disorder'' strength, $0\leq A\leq 1/2$). The limiting case $A=0$ corresponds to the ordered system
(the translation invariant Ising model) and $A=1/2$ corresponds to complete
disorder, where the spin lengths $\{\lambda _{k}^{\sigma }\}$ are uniformly
distributed over the interval $[0,1]$.

\begin{figure}[ph]
\centerline{\psfig{file=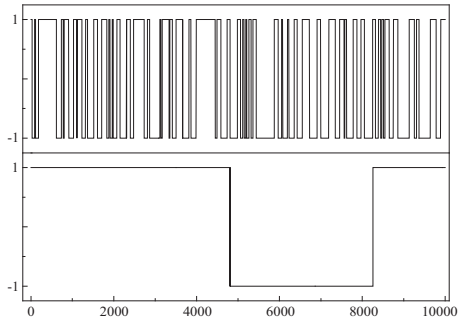,width=6.0cm}}
\vspace*{8pt}
\caption{The orientation of spins in the ground state of the systems for $N_e=10^4$
and two values of the disorder parameter $A$ of (\ref{s_k1}).
The top corresponds to $A=0.4$, the bottom to $A=0.2$. The
ground states have the domain structures and the domain concentration
increases with $A$.}
\label{Fig2}
\end{figure}

The examples of the ground state spin configurations for $N_{e}=10^4$ are
presented in Fig.~\ref{Fig2}.

We see that the ground state consists of ``domains'' \cite{PSS1} of blocks of
spins of the same sign, the domains concentration increasing rapidly with
the increase of disorder parameter $A$.

It follows from (\ref{lab}) and (\ref{HIs}) that interaction in our Ising
model is ferromagnetic although random. Thus, the ground state of (\ref{HIs}
) without the second term is ferromagnetic (collinear), i.e., with all the
spins either ``up'' or ``down''. Since, however, in our case the second term
(random field) is of the same order of magnitude as the first one
(interaction), an argument similar to that of the well known Larkin-Imry-Ma criterion \cite{Larkin,Imry-Ma,Ca:96} implies that the ferromagnetic ground state is unlikely for any $0\leq A\leq 1/2$. Thus, it is not completely unexpected that the ground state is not collinear. However, the Larkin-Imry-Ma argument does not suggest a detailed form of the genuine ground state, except that it has to be of a ``spin glass'' non-collinear type. On the other hand, our numerical method allows us to detect an explicit form of a ground state, having the domain structure.

We can also estimate the concentration $c_{dom}$ of the domain walls if $A$
is small enough. Indeed, the typical fluctuations of the pair interaction of
(\ref{Hub5}) (or of (\ref{HIs})) are
$$\delta \varepsilon \sim A^{2},$$
while energy of creation of a domain wall is
$$\varepsilon _{dom}\sim (\lambda_{k}^{+}+\lambda_{k}^{-})^{2}\sim 1,$$
if $A$ is not too close to 1. Thus, the domain is stable if its length
$l_{dom}$ satisfies the inequality
$$\delta \varepsilon \sqrt{l_{dom}}\lesssim \varepsilon _{dom}.$$

We obtain then \cite{PSS1,SK_2005}:
\begin{equation}
l_{dom}\sim A^{-4},  \label{l_dom}
\end{equation}
hence
\begin{equation}
c_{dom}=l_{dom}^{-1}\sim A^{4}.  \label{c_dom}
\end{equation}
In particular, it follows from (\ref{l_dom}) and (\ref{c_dom}) that the
ground state consists of a single ``ferromagnetic'' domain for $A=0$.

\begin{figure}[ph]
\centerline{\psfig{file=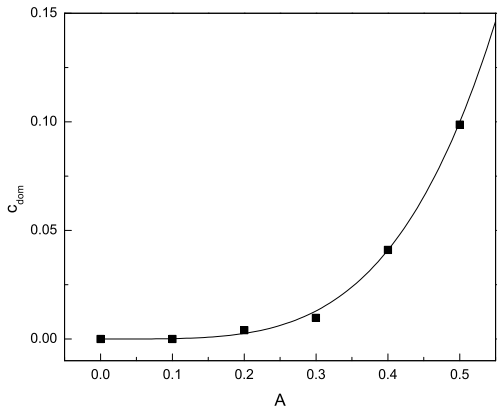,width=8.0cm}}
\vspace*{8pt}
\caption{The domain
concentration $c_{dom}$ as the function of disorder parameter $A$. Solid
boxes are the results of our numeric calculation, solid line is the fitting
by function $c_{dom}=(A/A_{0})^{d}$.}
\label{Fig3}
\end{figure}

We have calculated by the same method the domain concentration $c_{dom}$ for
a series of disorder parameter value $A$. The results are presented in 
Fig.~\ref{Fig3} (solid boxes).

The approximation by the function
$$c_{dom}=(A/A_{0})^{d}$$
is given by the solid line. The best fitting is for $A_{0}=0.88885$ and
$d=4.06031$ (quite close to those, obtained in
\cite{PSS1} and (\ref{c_dom})). Thus, we have a sufficiently good agreement between the numerical
data and the fitting curve for $A\leq 1/2$.

\section{Probing field.}

In this section we confirm our results of previous section on the form of
the ground states of model (\ref{HIs}) by probing its ground state by an
external field of special form.

Denote $H_{I}$ the r.h.s. of (\ref{HIs}) without $C$. Given a collection $%
b=\{b_{k}\}$, consider the perturbed Hamiltonian
\begin{equation}
H_{I}-\varepsilon \sum\limits_{k=1}^{N}b_{k}\sigma_{k}  \label{Hub5p}
\end{equation}
and the corresponding generalized magnetization
\begin{equation}
M(T,b)=-\frac{1}{N_{e}}\left. \frac{\partial F(T,b,\varepsilon )}{\partial
\varepsilon }\right\vert _{\varepsilon =0}=\frac{1}{N_{e}}%
\sum\limits_{k=1}^{N_{e}}b_{k}<\sigma _{k}>_{G},  \label{M}
\end{equation}
where $F(T,b,\varepsilon )$ is the free energy of (\ref{Hub5p}) and
$<...>_{G}$ denotes the corresponding Gibbs mean. By using the terminology of
spin glass theory (see e.g. \cite{Me-Co:87}), we can view (\ref{M}) as the
overlap between external field $\{b_{k}\}$ and the local magnetization
$\{<\sigma _{k}>_{G}\}$.

It follows from the r.h.s. of (\ref{M}) that the inequality
$$
|M(T,b)|\leq \left( \frac{1}{N_{e}}\sum\limits_{k=1}^{N_{e}}b_{k}^{2}\frac{1%
}{N_{e}}\sum\limits_{k=1}^{N_{e}}<\sigma _{k}>_{G}^{2}\right) ^{1/2}$$
holds for a generic $\{b_{k}\}$ and that it becomes the equality only for an
external field proportional to $\{<\sigma _{k}>_{G}\}$:
\begin{equation}
\overline{b}_{k}=a<\sigma _{k}>_{G},\;k=1,...,N_{e},  \label{hbar}
\end{equation}
where $a$ is a constant, i.e.,
\begin{equation}
M(T,\overline{b})=a\sum\limits_{k=1}^{N_{e}}<\sigma _{k}>_{G}^{2}.
\label{Mhbar}
\end{equation}
The constant $a$ can be chosen to be $1$ if we normalize $\{b_{k}\}$ by the
condition
\begin{equation}
\sum\limits_{k=1}^{N_{e}}b_{k}^{2}=1.  
\label{nob}
\end{equation}

Since we are interested in the ground state $\{\sigma _{k}^{GS}=\pm 1\}$ of
the Hamiltonian (\ref{HIs}), which we identify with the domain walls
configuration of the previous section, we put $T=0$ in the above formulas
and arrive to the following algorithm to detect $\{\sigma _{k}^{GS}\}$. Pick
a class of external fields containing $\{\sigma _{k}^{GS}\}$ and satisfying
(\ref{nob}) and vary $\{b_{k}\}$ over the class. The configuration
$\{\sigma_{k}^{GS}\}$ will be obtained as a maximizer of the generalized
magnetization (\ref{M}):
\begin{equation}
M(0,\{\sigma _{k}^{GS}\})=\frac{1}{N_{e}}\sum\limits_{k=1}^{N_{e}}(\sigma
_{k}^{GS})^{2}=1.  
\label{Mgs}
\end{equation}

In general, this will only prove that the corresponding maximizer is a local
minimum of the Hamiltonian (\ref{Hub5}) of the model, it is reasonable to believe
but the larger is the class the closer is the minimizer to the genuine ground state.

Let now $d_{1},d_{2},\ldots $ be the coordinates of domain walls of the spin
configuration found in the previous section ($<d_{m}-d_{m-1}>=l_{dom}$).
Consider the following class of random external (probing) fields:
\begin{equation}
b_{k}=\sigma _{k}^{GS}+f_{k},  \label{h_k}
\end{equation}
where
\begin{equation}
f_{k}=\left\{
\begin{array}{lr}
-2\sigma _{k}^{GS} & k\in (d_{m},d_{m}+\Delta _{m})\quad m=1,2,...,n, \\
0, & \mathrm{otherwise};
\end{array}
\right. ,  
\label{f_k}
\end{equation}

\begin{equation}
\Delta_m=(-1)^{\eta_m}[B\rho_m],  
\label{delta_m}
\end{equation}
$B \ge 0$ is a non-negative constant, $[\ldots ]$ denotes the integer part,
$\{\eta_m\}$ are the independent distributed random
variables, assuming ($0,1$) with probability $1/2$ and
$\rho_m$ are independent random variables with exponential distribution
$P(x)=B^{-1}\exp(-x/B)$.
Thus, the random variables $\{f_{k}\}$ provide generic fluctuations of the
positions of domain walls of the ground state configuration $\{\sigma _{k}^{GS}\}$. The
direction of the shift of the $m$-th domain wall is determined by $\eta _{m}$
and the amplitude of the shift is determined by $B$. In particular, the
probe field (\ref{h_k}) coincides with $\{\sigma _{k}^{GS}\}$ if $B=0$.

To find the free energy $F(T,b,\varepsilon )$ corresponding to (\ref{HIs}),
we use the transfer-matrix method. We set for $s_{k}^{\pm }=\pm \lambda
_{k}^{\pm },\;k=1,2,...,N_{e}$
\begin{equation}
{\hat{P}_{k}}(T,b)=\left(
\begin{array}{cc}
\exp \left( -\frac{(s_{k}^{+}-s_{k+1}^{+})^{2}-\varepsilon b_{k}s_{k}^{+}}{T}
\right) & \exp \left( -\frac{(s_{k}^{+}-s_{k+1}^{-})^{2}}{T}\right) \\
\exp \left( -\frac{(s_{k}^{-}-s_{k+1}^{+})^{2}}{T}\right) & \exp 
\left( -\frac{(s_{k}^{-}-s_{k+1}^{-})^{2}-\varepsilon b_{k}s_{k}^{-}}{T}\right)
\end{array}
\right),  
\label{P_k}
\end{equation}
assume the periodic boundary conditions ($\sigma _{N_{e}+1}=\sigma _{1}$)
and write
\begin{equation}
F(T,b,\varepsilon )=-T\log \left( \mathrm{Tr}\left[ \prod_{k=1}^{N_{e}}
{\hat{P}}_{k}(T,b)\right] \right),  
\label{F}
\end{equation}
where $\mathrm{Tr}$ denotes the trace of a $2\times 2$ matrix.

Using (\ref{P_k}) - (\ref{F}), one can calculate numerically the generalized
magnetization (\ref{M}) for the class (\ref{h_k}) -- (\ref{delta_m}) of
probing fields as the function of amplitude $B$ of fluctuations of the
domains walls:
\begin{equation}
M(T,b)\approx \frac{F(T,b,0)-F(T,b,\varepsilon )}{N_{e}\varepsilon }.
\label{M1}
\end{equation}

\begin{figure}[ph]
\centerline{\psfig{file=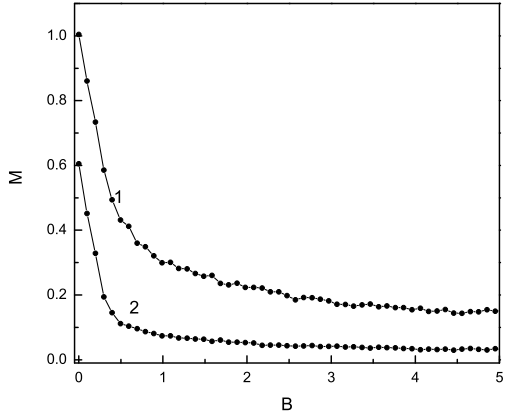,width=8.0cm}}
\vspace*{8pt}
\caption{Curve 1. The generalized magnetization $M(0,b)$ as the function of
$B$ for the fields (\protect\ref{h_k}) -- (\protect\ref{delta_m}
). Curve 2. The analogous curve for the probe fields with
$s_{k}^{GS}=\sigma_{k}^{GS}
\lambda_{k}^{\sigma_k^{GS}}$ instead of $\sigma_{k}^{GS}$ in the r.h.s. of (\ref{h_k}).}
\label{Fig4}
\end{figure}

Fig. \ref{Fig4} (curve 1) presents the generalized magnetization (\ref{M}) as
a functions $B$ for field (\ref{h_k}) and $T\rightarrow 0$.

We see that the magnetization is maximal for $B=0$, where the
fluctuations (\ref{f_k}) are absent, hence the probe field (\ref{h_k})
coincides with $\{\sigma_{k}^{GS}\}$. However, for $B>0$ the generalized
magnetization decays rather fast with the growth of $B$, i.e., it is a
rather sensitive characteristic of the proximity of the external field to
the maximizing one.

It is also worth noting that the if we replace the maximizing field $\sigma
_{k}^{GS}$ in (\ref{h_k}) by $s_{k}^{GS}=\sigma_{k}^{GS}
\lambda_{k}^{\sigma_k^{GS}}$ (see (\ref{sapm})) and compute again
numerically the corresponding generalized magnetization, we obtain a
curve (see curve 2 of Fig. \ref{Fig4}), which is quite similar to that of
curve 1 of Fig.\ref{Fig4}, except the magnitude of maximum, which is now $1-A$.

The magnitude can be explained as follows. In view of (\ref{sapm}) and (\ref%
{lab}) the maximum is
$$
\frac{1}{N_{e}}\sum\limits_{k=1}^{N_{e}} s_{k}^{GS}\sigma _{k}^{GS}=\frac{1}
{N_{e}}\sum\limits_{k=1}^{N_{e}} \alpha_{k}\sigma _{k}^{GS}+\frac{1}{N_{e}}
\sum\limits_{k=1}^{N_{e}} \beta_{k}.
$$
The second term on the right is $<\beta_k>=1-A$, if $N_e$ is large enough,
thus the first term has to vanish for $N_e >>1$ according our numerical
results. This can be viewed as a manifestation of a certain robustness of
our numerical algorithm to detect the ground state provided that the probe
field takes into account the sign structure of the ground state.

\section{Results and Discussion.}

The ground state of interacting particles on a disordered one-dimensional
host-lattice is studied using a new numerical method. It is shown that if the
concentration of particles is small enough, then even a weak disorder in
host-lattice site positions leads to the formation of the ``domains'' of
particles and to the breaking the long-range order pertinent to the
Generalized Wigner crystal in the absence of disorder. The nature of the
domains can be explained by using the Hubbard formula (\ref{x_k}) for
particle positions in the Generalized Wigner crystal. Indeed, in the case of
the translation invariant host-lattice the phase $\phi $ in the formula is
an arbitrary constant, just fixing the origin of the host-lattice. In the
disordered case each domains has its own phase, i.e., $\phi $ is a step
function with random steps and the mean domain length $l_{dom}$ is just the
mean length of steps.

It is also shown that the formula $l_{dom}\sim A^{-4}$
(\ref{l_dom}) is valid in a sufficiently wide range of disorder parameter
$A$: $0\leq A \leq 0.5$ (see Fig.~\ref{Fig3}).

The above results are then confirmed by studying the weak perturbations of
the Hamiltonian (\ref{HIs}) of the model by random external fields, which
``probes'' the domain structure of the ground state via random variations
(fluctuations) of the domain walls. It is shown that the generalized
magnetization per particle (\ref{M1}) corresponding to the field is maximal
if the amplitude of fluctuations is small and decays sufficiently fast with
the growth of the amplitude, i.e., that the magnetization is rather
sensitive to the proximity of the form of the field to that of the ground
state.


\begin{thebibliography}{0}
\bibitem{Mo:74} N. Mott, {\it Metal--Insulator Transitions} (Taylor and
Francis, London, 1974).

\bibitem{Slutskin_Gorelik} A. A. Slutskin and L. Yu. Gorelik, {\it Low Temp. Phys.} {\bf 19}, 852 (1993) [{\it Fiz .Niz. Temp.} {\bf 19}, 1199 (1993)].

\bibitem{Hubbard1} J. Hubbard, {\it Phys. Rev. B} {\bf 17}, 494 (1978).

\bibitem{nano_dot} E. Y. Andrei, {\it 2D Electron Systems on Helium and Other
Substrates} (Kluwer, New York, 1997).
\bibitem{nano_grain} H. Nejoh, M. Aono, {\it Appl. Phys. Lett.} {\bf 64}, 2803 (1995). 
\bibitem{Bak} P. Bak, R. Bruinsma,  {\it Phys. Rev. Lett.} {\bf 49}, 249 (1982).
\bibitem{Sinay} S. E. Burkov, Ya. G. Sinai, {\it Russ. Math. Surv.} {\bf 38}, 235 (1983).
\bibitem{PRB96} V. V. Slavin, A. A. Slutskin, {\it Phys. Rev. B} {\bf 54}, 8095  (1996).
\bibitem{PRB00} A. A. Slutskin, V. V. Slavin, and H. A. Kovtun,
{\it Phys. Rev. B} {\bf 61}, 14184 (2000).
\bibitem{Larkin} A. I. Larkin, {\it Sov. Phys. - JETP} {\bf 31}, 784 (1970).
\bibitem{Imry-Ma} Y. Imry, S. K. Ma, {\it Phys.Rev.Let.} {\bf 35}, 1399 (1975). 

\bibitem{Frattini1} S. Fratini, B. Valenzuela and D. Baeriswyl, {\it
arXiv:0209518v1}.

\bibitem{Frattini2} S. Fratini, B. Valenzuela and D. Baeriswyl, {\it arXiv:0302020v2}.

\bibitem{Leib} E. H. Lieb, F. Y. Wu, {\it Physica A} {\bf 321}, 1 (2003).

\bibitem{Jedrzejewski} J. Jedrzejewski, J. Miekisz, {\it arXiv:9903163}.

\bibitem{Ca-Co:08} A. Campa, T. Dauxois, S. Ruffo, {\it Physics Reports} {\bf 480}, 57 (2009).

\bibitem{Da-Co:09} T. Dauxois, S. Ruffo, L. Cugliandolo (Editors),
{\it Long-Range Interacting Systems, Lecture Notes of the Les Houches Summer
School 2008} (Oxford University Press, Oxford 2009).

\bibitem{Mu-Co:10} F. Bouchet, S. Gupta, D. Mukamel, {\it Physica A} {\bf 389}, 389 (2010).

\bibitem{PSS1} V. V. Slavin, {\it Phys. Stat. Sol. (b)} {\bf 241}, 2928 (2004).

\bibitem{LGP:88} I. M. Lifshitz, S. A. Gredeskul, L. A. Pastur,
{\it Introduction to the Theory of Disordered Systems} (Wiley, NY, 1988).
\bibitem{Gr:12} A. Grechnev, {\it arXiv:1212.2320}.
\bibitem{Du:12} Y. I. Dublenych, {\it Phys. Rev. Lett.} {\bf 109}, 167202 (2012).

\bibitem{Ca:96} J. Cardy, {\it Scaling and Renormalization in Statistical
Physics} (Cambridge University Press, 1996).

\bibitem{SK_2005} A. A. Slutskin and H. A. Kovtun, {\it Low Temp.Phys.} {\bf 31} 594 (2005) [{\it Fiz. Niz. Temp.} {\bf 31}, 784 (2005)].

\bibitem{Me-Co:87} M. Mezard, G. Parisi, M. A. Virasoro, {\it Spin Glass and
Beyond} (World Scientific, Singapore, 1987).
\end{thebibliography}
\end{document}